\documentclass[12pt]{article}
\usepackage[english]{babel}
\usepackage{graphicx}
\usepackage{amsmath}
\usepackage{endnotes}
\usepackage{braket}
\usepackage{amssymb}
\usepackage[makeroom]{cancel}
\usepackage{accents}
\usepackage[colorlinks]{hyperref}
\usepackage{makeidx}
\usepackage{setspace}
\usepackage[utf8]{inputenc}
\usepackage{hyperref}
\usepackage{braket}

\usepackage[colorlinks]{hyperref}
\usepackage{color}

\definecolor{navyblue}{rgb}{0.0, 0.0, 0.5}
\definecolor{darkgreen}{RGB}{0,80,0}
\definecolor{darkgray}{RGB}{0,80,0}
\definecolor{darkred}{RGB}{80,0,0}
\definecolor{venetianred}{rgb}{0.78, 0.03, 0.08}
\definecolor{violet(colorwheel)}{rgb}{0.5, 0.0, 1.0}
\definecolor{amber}{rgb}{1.0, 0.75, 0.0}
\definecolor{marron}{RGB}{60,30,10}
\definecolor{darkblue}{RGB}{0,0,80}
\definecolor{lightblue}{RGB}{80,80,80}

\definecolor{shadecolor}{rgb}{0.97,0.97,0.97}
\definecolor{bubblegum}{rgb}{0.99, 0.76, 0.8}

\usepackage[top=3cm, bottom=3cm, left=3cm, right=3cm]{geometry}
\begin{document}

\begin{center}
\Large{\bf{Is Time Travel Too Strange to Be Possible?}}   \\
 \vspace{7 mm}  \setstretch{0,100}{\large {\bf{Determinism and Indeterminism on Closed Timelike Curves}}}\\
\vspace{1,2cm}
\large{Ruward A. Mulder and Dennis Dieks}\\
\vspace{0.8cm}
\normalsize{\emph{History and Philosophy of Science, Utrecht University, \\Utrecht, The Netherlands}}
\end{center}
\vspace{0.4cm}

\setstretch{1.0}
\begin{abstract}
\footnotesize Notoriously, the Einstein equations of general relativity have solutions in which closed timelike curves (CTCs) occur. On these curves time loops back onto itself, which has exotic consequences: for example, traveling back into one's own past becomes possible. However, in order to make time travel stories consistent constraints have to be satisfied, which prevents seemingly ordinary and plausible processes from occurring. This, and several other ``unphysical'' features, have motivated many authors to exclude solutions with CTCs from consideration, e.g. by conjecturing a chronology protection law.

In this contribution we shall investigate the nature of one particular class of exotic consequences of CTCs, namely those involving  unexpected cases of indeterminism or determinism. \emph{Indeterminism} arises even against the backdrop of the usual deterministic physical theories when CTCs do not cross spacelike hypersurfaces outside of a limited CTC-region---such hypersurfaces fail to be Cauchy surfaces.  We shall compare this \textit{CTC-indeterminism} with four other types of indeterminism that have been discussed in the philosophy of physics literature: quantum indeterminism, the indeterminism of the hole argument, non-uniqueness of solutions of differential equations (as in Norton's dome) and lack of predictability due to insufficient data.

By contrast, a certain kind of \emph{determinism} appears to arise when an indeterministic theory is applied on a CTC: things cannot be different from what they already were. Again we shall make comparisons, this time with other cases of determination in physics.

We shall argue that on further consideration both this indeterminism and determinism on CTCs turn out to possess analogues in other, familiar areas of physics. CTC-indeterminism is close to the epistemological indeterminism we know from statistical physics, while the ``fixedness'' typical of CTC-determinism is pervasive in physics. CTC-determinism and CTC-indeterminism therefore do not provide incontrovertible grounds for rejecting CTCs as conceptually inadmissible.
\end{abstract}

\pagenumbering{gobble}
 \clearpage
\pagenumbering{arabic}
\section{Introduction}
There have been extensive discussions in the philosophical and physical literature of the last couple of decades about the possibilities of time travel: the existence of solutions of the Einstein equations in which closed time-like curves (CTCs) occur has endowed the science-fictional character of the subject with a certain amount of scientific respectability. That the Einstein equations of general relativity do not exclude the existence of CTCs is easy to see. The Einstein equations impose \emph{local} conditions on spacetime: the local curvature properties must stand in a definite relation to the local energy and momentum of the matter fields. As long as these local conditions remain satisfied, the global topology of the spacetime may vary. Now, one particular solution of the Einstein equations is Minkowski spacetime, in which the curvature vanishes everywhere (Minkowski spacetime is flat) and in which there is no matter (all components of the energy-momentum tensor are zero). From this Minkowski spacetime we can build a new solution of the Einstein equations, with a different topology, by the simple operation of identifying two spacelike hypersurfaces (one ``in the future'', and one ``in the past''). Concretely, we cut a strip out of Minkowski spacetime and glue the upper and lower ends together. The cylindrical spacetime that results (a strip of Minkowski spacetime rolled up in the time direction) features CTCs: timelike worldlines going straight up in the time direction return to their exact starting points.

It is helpful to keep this simple example of CTCs provided by rolled-up Minkowski spacetime in mind during the discussion of strange features of CTCs below. However, it should not be thought that CTCs only arise by (arguably artificial) cut-and-paste constructions: there are quite a number of solutions of the Einstein equations known in which apparently plausible matter distributions give rise to CTCs.  The most famous solution with CTCs was found by Kurt G\"odel (G\"odel, 1949, \cite{Godel}). The G\"odel solution describes a non-expanding, rotating universe with a large cosmological constant. This G\"odel world is conceptually important, even though it has properties that conflict with what we empirically find in our universe. A solution of the Einstein equations that may be more relevant for our own world is that of a part of spacetime with a spinning black hole (described by the Kerr metric), in which a region with CTCs exists, the \textit{ergosphere} (see, for example, Carroll, 2014, \cite{Carroll}, p.\ 261.). There are also other, less drastic mass-energy distributions that feature CTCs, such as the first metric with CTC-properties to be discovered: van Stockum's rotating cylinder of dust particles (van Stockum, 1937, \cite{Stockum}).

It should be added that it is unclear whether the existence of these and similar solutions has the implication that there are possibilities for building a ``time machine''. The construction of a manageable time machine would require the creation of a singularity-free compact spacetime region in which CTCs occur, and results by Hawking and others indicate that this cannot be realized within classical general relativity because of energy-conditions that have to be satisfied by the matter fields (Hawking, 1992, \cite{Hawking}). In a quantum theory of general relativity there might be more room for time machines, but this remains speculative in view of the absence of a full quantum gravity theory. However, we are not concerned with the question whether it is possible to construct a useful time machine, but rather with the conceptual status \emph{per se} of CTCs, in relativistic worlds in general.

It is immediately clear that CTCs give rise to causal oddities. If someone could go back into his own past, even if not in our universe but in another physically possible world, it seems that he could change things there in such a way that logical inconsistencies result. The notorious example is the grandfather paradox: if a time traveler arrives at a point in his past at which his grandfather is still an infant, he could decide to kill the child. But if this were to be successful, it would obviously conflict with the very fact of the time traveler's own existence. So in order to have a consistent history on a CTC, certain consistency conditions have to be fulfilled. These conditions take the form of restrictions on what can happen. Although in the case of the grandfather paradox it is plausible at first sight that the time traveler can do whatever is within his capabilities when meeting his grandfather, he nevertheless must be constrained in his actions. This points into the direction of a \emph{determinism} that is stricter than what we are used to in physics; this is one of the points to be further discussed below.

In contrast, there are also cases in which the presence of CTCs appears to lead to a weakening of usual notions of determinism. Think of a spacetime that is globally Minkowski-like, but in which there is a finite region containing CTCs that are causally closed within themselves, so that nothing happens on them which has a cause external to the CTC. Closed worldlines of inertially moving particles would be an example. Since there is no causal relation to anything outside such worldlines, the processes that take place in the CTC region cannot be predicted from outside that region. This lack of predictability points into the direction of \emph{indeterminism}, the other issue to be discussed in more detail below.

The threats posed by logical inconsistency, causal anomalies and other strange features have sometimes been adduced to declare the possibility of CTCs ``unphysical''. CTCs should perhaps be ruled out by a ``cosmic censorship principle'' (Penrose, 1968, \cite{Penrose}) or a ``chronology protection principle'' (Hawking, 1992, \cite{Hawking}). The motivation behind conjecturing that such a principle is at work is that the alternative, with its CTC extravagances, goes against the very nature of physics: such features occur nowhere else in existing physical theory or practice. This then is taken to justify the inductive conclusion that they cannot happen at all.

In this paper we shall critically analyze this argument for the impossibility of cases of ``unphysical determinism and indeterminism.'' As we shall argue, it is not accurate to say that the ``causal anomalies'' associated with CTCs form a category of their own. Indeed, there exist similar cases in standard physics, and these cases are not considered to be exotic or even strange. Therefore, we shall conclude, such \textit{a priori} objections are not insuperable and hence at least part of the motivation for the dismissal of CTCs falls away.

\section{Closed Timelike Curves and Logical Consistency}
\subsection{A Toy Model: Deutsch-Politzer Spacetime}\label{DP}
\begin{figure}
\centering
\includegraphics[scale=0.7]{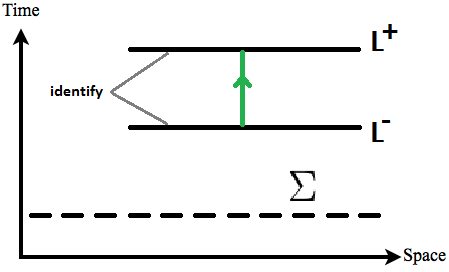}
\caption{Deutsch-Politzer spacetime, picture is based on (Arntzenius $\&$ Maudlin, 2013). $L^+$ and $L^-$ represent the `cuts' in Minkowski spacetime; the lower side of $L^+$ is identified with the upper side of $L^-$, while the lower side of $L^-$ is identified with the upper side of $L^+$. This can be visualized as a ``handle" on Minkowski spacetime, which particles may enter and leave again (in contrast with the rolled-up cylinder). $\Sigma$ is a spacelike hypersurface outside of the CTC-region; it is not a Cauchy surface. The green line represents a particle at rest, with a closed timelike worldline.} \label{fig:DPST}
\end{figure}

A spacetime that is a bit more complicated than the rolled-up version of Minkowski spacetime that we considered in the Introduction, and which is suitable for  illustrating our arguments, is the so-called Deutsch-Politzer spacetime (Deutsch, 1991, \cite{Deutsch}; Politzer, 1992, \cite{Politzer}). It can be realized by making two cuts in flat Minkowski spacetime, as indicated in Figure \ref{fig:DPST}---the points in these cuts are removed from the manifold. The two inner edges of the cuts are subsequently identified (glued together); the same happens with the two outer edges. When a particle hits $L^-$ from below it will travel onward from the upper side of $L^+$. The other way around, a particle hitting the lower side of $L^+$ will reappear at the upper side of $L^-$. This creates a region where CTCs can occur: vertical lines, which represent particles at rest (in the depicted frame of reference) that would loop back onto themselves, as depicted by the particle worldline in the figure.

This cut-and-paste operation results in Minkowski spacetime with a kind of ``handle'', the latter having the internal topological properties of a rolled-up strip of Minkowski spacetime. The spacetime region of the handle can be entered from the rest of spacetime, namely from the two sides, on the left and on the right, where the handle merges with the surrounding spacetime. One can think of this spacetime as global Minkowski spacetime, which at two singular points makes contact with a rolled-up strip of Minkowski spacetime.\footnote{It has been shown that for the Deutsch-Politzer spacetime it is not possible to smooth out the metric such that we would obtain a global nonsingular asympotically flat Lorentzian metric (Chamblin et al., 1994, \cite{Chamblin}). This is because the end points of the ``cuts'' are singularities: as judged from the surrounding Minkowski spacetime, there is a finite spacetime interval between the lower and upper end points on both sides, but seen from the inside these points are identical. These two singularities raise questions about the empirical plausibility of Deutsch-Politzer spacetime. Nevertheless, consideration of this spacetime is helpful as it makes visualization of finite CTC-regions possible;  our conclusions will not depend on a commitment to this or another specific spacetime.}

\begin{figure}
\centering
\includegraphics[scale=0.7]{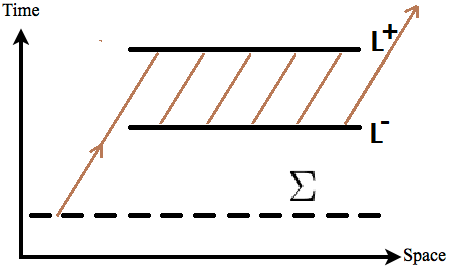}
\caption{ A particle traveling through Deutsch-Politzer spacetime. The particle enters the CTC-region with a certain velocity, and after hitting $L^+$ five times, it continues its path in the outer Minkowskian region. Picture is based on (Arntzenius and Maudlin, 2013).} \label{fig:DPST2}
\end{figure}

Because worldlines inside the handle do not cross the spacelike hypersurface $\Sigma$ (see Figure \ref{fig:DPST}), $\Sigma$ is not a Cauchy surface: specification of the physical state on $\Sigma$ does not fix the physical state on the whole manifold. In particular, there is no information available on $\Sigma$ about what happens inside the handle. There is consequently a ``Cauchy horizon'', which limits the part of spacetime that can be predicted from $\Sigma$. The unpredictable part comprises the handle, and also the two future lightcones with their apexes in the two singular endpoints of the handle. In Figure \ref{fig:DPST} these two singularities are represented by the two ends (on the left and right, respectively) of the lower cut. Since the singularities are not part of the manifold, they cannot contain initial conditions for worldlines that ``come out of them'' (this is a figurative way of speaking, since the singular points are not in the manifold and therefore not on any curve in the manifold---the past parts of the worldlines in question asymptotically approach one of the singularities). These worldlines do not have a starting point, and can in this respect be compared to worldlines coming in from infinity. There is no origin of such worldlines whose properties determine their nature or number. Clearly therefore, data available on $\Sigma$ cannot fix what is beyond the Cauchy horizon.

\subsection{Consistency Constraints: The Grandfather `Paradox'}\label{CC}
In the region of the handle in Deutsch-Politzer spacetime CTCs occur. These CTCs raise questions about the consistency of histories and the determination of events, as already mentioned in the Introduction.

For example, a person whose worldline is one of these CTCs will return into his own past, where he must find himself in exactly the same state as on earlier occasions. This uniqueness is simply a logical consequence of the demand that histories on the CTC must be unambiguous: there has to be exactly one physical state of affairs at each point of the CTC. This uniqueness gives rise to a consistency constraint: We can only have solutions that are consistent in the sense that they consist of well-defined unambiguous events along the CTC. This is a logical truism and as such a harmless requirement. Nevertheless, in the present context these consistency constraints introduce a restriction on possible histories that seems counter-intuitive and leads to what Smeenk and W\"utrich (2011, \cite{Smeenk}, p.\ 7) call \textit{modal paradoxes}, of which the grandfather paradox is a concrete instance.

In the grandfather paradox our time traveler, living on a CTC, goes back in time and meets his grandfather. Obviously, everything will have to happen in exactly the same way as recorded in history. In particular, it cannot happen, on pain of logical inconsistency, that the grandson undertakes actions that make his own birth impossible. The paradox, as usually formulated, is that this seems to take away some of the powers of the grandson: surely, one is apt to argue, he is \emph{able} to kill his grandfather. So how could it be that he cannot in fact do so? Stephen Hawking has argued that such paradoxes threaten time travel with ``great logical problems" and that we should hope for ``a Chronology Protection Law, to prevent people going back, and killing our parents" (Hawking, 1999, \cite{Hawking2}). Such a novel Law is not needed, though: as we shall argue the trivial constraint that everything is well defined and consistent will do the job.

Before discussing this further, we should mention that a more complicated solution of the paradox was suggested by David Deutsch in his quantum model of time travel. Here, physical systems can traverse so-called Deutsch-CTCs that go from one world to another in Everettian many-worlds quantum mechanics (Deutsch and Lockwood, 1994, \cite{Deutsch2}). This model connects events in different worlds, and the model therefore basically invokes multiple time dimensions---as noted, e.g., in (Dunlap, 2016, \cite{Dunlap}). It is true that multidimensional time offers a way out of  the grandfather paradox, since the fact that a grandson killed his grandfather in a world with time $t_2$ does not contradict that his grandfather stays alive in a world with time $t_1$, in which the grandson was born (cf.\ Dainton, 2010, \cite{Dainton}, p.\ 123). However, this response of invoking multiple time dimensions seems artificial in solving issues with time travel and, moreover, fails to address the original paradox, in which the time traveler goes back to his own past in his own time in his own world. It is only this original paradox that we shall consider here.

An important remark concerning the original paradox, which takes away some of the puzzlement, was made by David Lewis. As Lewis points out, in the formulation of the paradox there is an ambiguity in the use of `can' and `being able' (Lewis, 1976, \cite{Lewis}). Of course, the time traveling grandson \emph{is able} to kill his grandfather in the sense that he knows how to use a firearm, has the required muscular strength, training, and so on. But not everything that `can' be done in this sense will \textit{actually} be done---we do not at all need to consider time travel to recognize this and to realize that there is no contradiction here. In fact, it is a general truth, also in universes without CTCs, that only one act among all the acts that one is able to do will actually be done. In the case of a CTC, the grandson can accordingly be assumed to be able to kill his grandfather, in the sense of possessing the required means and capacities: he `can' shoot. But at the same time it is impossible that he will actually do so: if he did, he would contradict the historical record. Accordingly, no immediate contradiction between `can' and `cannot' arises. According to Lewis the paradox is therefore only apparent; there is an ambiguity in the word `can', which is not always visible, but which is highlighted in some cases---and it is highlighted very prominently in time travel situations. That the time travel story does not sit well with common sense is simply because we are not used to CTC-like situations.

Lewis' analysis is correct in our opinion, but its emphasis on human acts and capabilities invites questions, e.g.\ about the nature of volition and human powers, that distract from the physical aspects of the problem. We shall therefore discuss variations on the paradox that only involve physics. In section \ref{DtoI} we shall consider particles that obey \emph{deterministic} dynamical laws; in section \ref{ItoD} we shall consider an \emph{indeterministic} physical process (like radio-active decay, governed by quantum mechanics). As it will turn out, application of a deterministic theory to a spacetime in which there is a CTC-region leads to a particular kind of indeterminism. In section \ref{com}, we shall compare this CTC-indeterminism to other forms of indeterminism we know from physics. By contrast, in the case of an indeterministic quantum process, grandfather-paradox-like reasoning on a CTC will lead us to determinism---at first sight in conflict with what the standard interpretation of quantum mechanics tells us about the essentially indeterministic nature of the theory.

The latter result, the appearance of determinism in an initially indeterministic context, may perhaps be expected since we have already seen that consistency on CTCs reduces possibilities. That there is also a counterpart to this, namely the appearance of \emph{indeterminism}, has also already been indicated, in the previous subsection: beyond the Cauchy horizon there are worldlines that cannot be fully determined from $\Sigma$. This can be used to construct examples of indeterminism, even if the local physical laws are deterministic.

\section{From Determinism to Indeterminism} \label{DtoI}
In our construction we shall use the Deutsch-Politzer spacetime explained in section \ref{DP}. In Figure \ref{fig:DPST}, the chronology violating region (i.e., the region where CTCs occur) is located to the future of a spacelike hyperplane $\Sigma$. No worldlines from this region cross $\Sigma$, and in general no wordlines cross $\Sigma$ more than once. Everything looks  therefore ``normal'' on $\Sigma$, just as on an arbitrary hyperplane in Minkowski spacetime. However, the initial value problem on $\Sigma$ is not well-posed because $\Sigma$ does not qualify as a Cauchy surface. Indeed, the standard definition of a (global) Cauchy surface $\Sigma$ in a spacetime manifold $\mathcal{M}$ is a surface that is intersected exactly once by \emph{every} non-spacelike curve in $\mathcal{M}$. It is understandable that initial conditions on such a surface $\Sigma$ determine all events in $\mathcal{M}$ if the applicable laws of physics are locally deterministic, since the physical state on $\Sigma$ is propagated by these laws along the non-spacelike curves of $\mathcal{M}$. In the case of a Cauchy surface these curves fill the entire spacetime. But in the case of Figure \ref{fig:DPST} there clearly are worldlines that do not intersect $\Sigma$ and about whose behavior no information is available on $\Sigma$. Consequently, the state on $\Sigma$ does not contain enough information to fix the entire global state of $\mathcal{M}$.

One could say that the CTC-region to some extent forms a world in itself. It is true that particles can enter the CTC-region from outside, such as in Figure \ref{fig:DPST2}, and it is true that this can be predicted from the initial conditions on $\Sigma$. However, one may add an arbitrary number of undetermined worldlines with a particle on it, beyond the Cauchy horizon. This will of course lead to different global states of $\mathcal{M}$, with a different total mass and energy. Therefore, associated with any initial state on $\Sigma$ is an infinitude of global states of $\mathcal{M}$. This seems a clear case of indeterminism, which will typically occur in spacetimes in which isolated chronology violating regions occur. We shall christen this kind of indeterminism ``CTC-indeterminism.''

\section{CTC-Indeterminism Among Other Varieties of Indeterminism} \label{com}

In this section we shall compare the indeterminism that we have seen to arise when CTCs are present with other cases of indeterminism in physics. If the argument that CTCs can be dismissed because of their exotic and unphysical features is to work, the indeterminism that arises here should be in a category of its own, different from cases we encounter elsewhere in physics. In order to see whether this is in fact so, we shall successively review the indeterminism of Norton's Dome, the indeterminism of quantum mechanics, that of the hole argument, and finally the indeterminism of statistical physics.

\subsection{Norton's Dome}
Norton's Dome is an example in which Newton's second law of motion fails to have one unique solution (Norton, 2008, \cite{Norton1}). In this case the mathematical condition for the relevant differential equation to have a unique solution, the so-called Lipshitz continuity condition, is not satisfied.\footnote{The Lipshitz condition is the demand that the slope of the force function does not become too large. Specifically, a function $F$ satisfies the condition within a certain domain $D$ iff there is a constant $K>0$ such that $|F(x)-F(y)|\leq K|x-y|$. For a detailed discussion, see (Fletcher, 2012, \cite{Fletcher}).}

\begin{figure}[h!]
\centering
\includegraphics[scale=1]{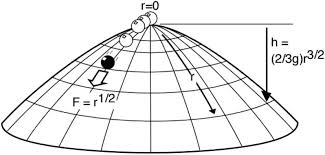}
\caption{Norton's Dome from (Norton, 2008, \cite{Norton1}). The marble at the top of this dome is initially resting, but will spontaneously---that is, unpredictably---roll down the surface, as shown by the solution \ref{solution}. } \label{fig:dome}
\end{figure}

The set-up is as follows. Consider a dome-like surface, as shown in Figure \ref{fig:dome}, in a uniform gravitational field and with a particle of mass $m$ that can move on it. The height $h(r)$ of the surface of the dome as a function of the radial coordinate $r$, is given by
\begin{equation}
h(r)=\frac{2\alpha}{3g}r^{3/2},
\end{equation}
where $g$ is the gravitational acceleration, $\alpha$ a proportionality constant.  \\

The net force on a particle at the surface will be tangentially directed and is hence given by
\begin{equation}
F=mg\sin\phi=mg\frac{dh}{dr}= \alpha \, m  r^{1/2},
\end{equation}
with $\phi$ the angle between the tangent and the horizontal direction. Newton's second law takes the form of the differential equation
\begin{equation} \label{2L}
\frac{d^2r(t)}{dt^2}=\alpha \,  r^{1/2}.
\end{equation}
 If the initial situation is a particle at rest at the top of the dome an obvious solution of Eq. \ref{2L} is $r=0$, $\forall t$. However, there are also other solutions with the same initial condition, namely\footnote{This is a solution since
\begin{equation}
\frac{d^2}{dt^2}\frac{\alpha^2}{144}(t-T)^4=\frac{\alpha^2}{12}(t-T)^2=\alpha\sqrt{\frac{\alpha^2}{144}[(t-T)^4}=\alpha r^{1/2}, \nonumber 
\end{equation} satisfying Eq.\ \ref{2L}.}
\begin{equation} \label{solution}
r(t) =
\left\{
	\begin{array}{ll}
		0  & \mbox{if } t<T, \\
		\frac{\alpha^2}{144}[(t-T)]^4	  & \mbox{if } t>T, \\
	\end{array}
	\right.
\end{equation}
for an arbitrary value of $T$. Because of the arbitrariness of $T$, there is an infinity of possible solutions to this differential equation. Eq.\ \ref{solution} describes a particle initially at the origin, which starts rolling off the surface after the time $T$ has elapsed. Because the initial conditions do not fix one unique motion, determinism fails.

When comparing this indeterminism to our case of CTC-indeterminism, we see that there are various differences. In the dome case, all possibly relevant initial data have been specified, but the differential equation is not able to produce one unique solution from them because the Lipshitz condition is violated. In the CTC case the initial conditions and forces on $\Sigma$ do produce unique worldlines---Lipshitz conditions are everywhere assumed to be satisfied. But these uniquely determined worldlines departing from $\Sigma$ do not determine what goes on in the chronology violating region. To know how many particles find themselves in the handle of Figure \ref{fig:DPST} we need \emph{more initial data}, and these data are not available on $\Sigma$. We can conclude that CTC-indeterminism is not connected to the violation of a Lipshitz condition. CTC-indeterminism is therefore essentially different from the indeterminism in the dome case.

\subsection{Quantum Indeterminism}
The indeterminism of quantum mechanics is given a precise form by the Born rule, which states that the square of the amplitude of a particular term in the quantum state (written down as a superposition in some basis) yields the \emph{probability} for finding a measurement outcome corresponding to that term. According to the standard interpretation of quantum mechanics this probability is fundamental: it is not possible to refine the description by adding parameters to the wave function, in such a way that the predictions become more precise than allowed by the Born rule. In other words, even if the state of a physical system is completely specified, the theory only provides us with \emph{probabilities} for a \emph{range} of possible measurement outcome---the theory is thus indeterministic.

CTC-indeterminism and the indeterminism resulting from the Born rule have in common that in both cases a unique prediction of measurable quantities cannot be fixed by specifying all initial conditions at a given time (that is, say, on some spatial hypersurface). However, in the case of CTC-indeterminism the applicable equations do not tell us anything at all about probabilities for the different possibilities. One may draw an arbitrary amount of additional causally closed particle worldlines inside the CTC-region, which corresponds to an infinitude of possibilities, but the equations do not assign any chances to them---the equations do not speak about probabilities at all. This is an essential difference with quantum theory, in which the laws possess a probabilistic interpretation from the outset. Put differently, CTC-indeterminism does not come from the probabilistic character of the applicable theory, whereas quantum indeterminism does.

In section \ref{ItoD} we will come back to the status of probabilistic theories when applied to CTC-regions. For now, it suffices to observe that CTC-indeterminism is not similar to quantum indeterminism. In the latter the specification of probability values is essential, whereas in the former probabilities never enter the discussion.

\subsection{The Hole Argument}
The hole argument, basically devised by Einstein in 1913, makes use of the background independence of general relativity, which ensures that all different coordinate systems perform \emph{a priori} equally well when used to express the laws of the theory---there is no \emph{a priori} given spacetime geometry which could define a privileged frame of reference and coordinates adapted to it (Dieks, 2006, \cite{Dieks1}). The modern form of the argument was developed by (Stachel, 2014, \cite{Stachel}) and (Norton and Earman, 1987, \cite{EarmanNorton}), and extensively used in discussions about substantivalism versus relationism with respect to spacetime.

\begin{figure}
\centering
\includegraphics[scale=0.6]{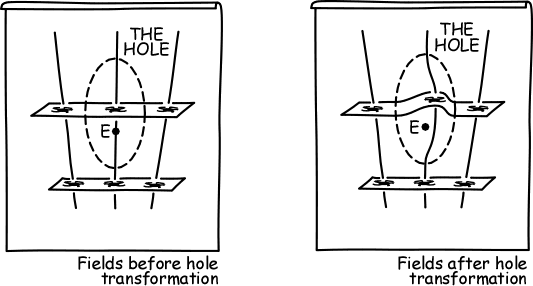}
\caption{The hole argument as presented by Norton (Norton, 2015, \cite{Norton2}). The invariant aspects of metric and matter fields of particles (here, even entire galaxies) and their distribution in spacetime are the same, although the events are differently placed in the manifold. The transformation from one situation to the other is accomplished by a `hole transformation' (a diffeomorphism) in the area indicated by the dotted line. The question then is: does the galaxy pass through spacetime point $E$ or not? }
\label{fig:hole}
\end{figure}

We are concerned with only one possible implication of the hole argument, namely that it proves general relativity to be indeterministic. The argument is as follows. The spacetime of general relativity in a concrete situation contains a labeled set of events placed in a manifold, plus a metric field specifying temporal and spatial relations between the events. Not all the degrees of freedom in the theory are physical, some of them are concerned with how the events are placed in the manifold. Because of the background independence this manifold does not possess a pre-given geometry, so that one can perform a gauge transformation (an active coordinate transformation) on this distribution of events---see Figure \ref{fig:hole}. All observable properties can be reduced to combinations of relativistic invariants and are left intact by such a transformation. If the transformation constitutes a real change in the world (as the spacetime substantivalist would maintain) then this real change---a different distribution of matter and energy over the manifold---cannot be dealt with by the theory: the initial and boundary conditions do not fix which one of the different possible distributions will be realized. This is because any two distributions of metric and matter agree on all observational properties, which are the only properties predicted by the dynamical laws of the theory. This amounts to indeterminism.

Compared to CTC-indeterminism, this type of indeterminism shares the characteristic that it leaves open an infinite range of possibilities, and does not assign any probabilities to these different options. Moreover, like CTC-indeterminism, hole indeterminism does not depend on violations of Lipshitz conditions: all differential equations are perfectly well behaved and have unique solutions in terms of invariant quantities. It is only because of general covariance, relating to the absence of a fixed spacetime background, that equally valid spacetime \emph{representations} of the physical situation (characterized by invariant quantities) arise---these representations relate to each other via diffeomorphisms.

The latter observation marks an essential difference between CTC-indeterminism and hole indeterminism. In the case of hole indeterminism all diffeomorphically related possibilities feature exactly the same values of all physical quantities like energy, mass, etc. For this reason, it may be argued that these different solutions are actually physically identical---as relationists do.  This is completely unlike the situation in CTC-indeterminism: here the possibilities are uncontroversially physically different, distinguishable as they are on the basis of the numbers of particles inside the CTC-region, the amounts of mass and energy, and so on. So CTC-indeterminism and hole indeterminism belong to very different categories.

\subsection{Lack of Knowledge}
When the state of a physical system is incompletely specified, it is to be expected that its future behavior cannot be fully predicted, even if the applicable laws are deterministic and if the Lipshitz conditions are satisfied. For example, if we only know the initial positions and velocities of a restricted number of particles, or only possess estimates for these quantities, Newton's equations will not enable us to accurately predict the future state of a many-particle system. More than one final state will be compatible with the initial data.

An example of such a situation in physical practice is the micro-description of systems characterized by macro-quantities, as in statistical physics. Statistical physics applies when the number of degrees of freedom of a system becomes too large to be practically tractable, as in the case of a macroscopic amount of a gas or the modeling of noise-effects in electronic devices. Although the laws of classical mechanics fix a unique evolution given the initial state, it is practically impossible to ascertain all individual initial momentum and position values. Statistical physics deals with this by considering ensembles of \textit{possible} microstates. 

The crucial point is that the macrostate underdetermines the microstate: there are many microstates that would give rise to the same macrostate. Hence, statistical physics can be seen as quantifying our ignorance about the microphysical state. This leads us to a ``pragmatic'' type of indeterminism: in the absence of a full set of initial conditions, the future lies open---in the epistemic sense.

Clearly, this underdetermination is only to be expected when information is missing. The associated indeterminism is consequently harmless from a fundamental point of view: determinism can be restored by making the description more complete. In the ontological sense, only one microstate is actually realized, and this state has a fully deterministic evolution.  In the next section, we argue that this non-fundamental indeterminism is very similar to CTC-indeterminism.

\subsection{CTC-Indeterminism as Epistemic and Harmless}
CTC-indeterminism comes about as the result of the existence of Cauchy horizons, which restricts the amount of information available on  spacelike hypersurfaces---what is beyond the horizon is hidden from view. This indicates that CTC-indeterminism is of the same kind as the epistemically founded indeterminism just discussed, and hence of the harmless kind. This suggestion is supported by our earlier observation that CTC-indeterminism is fundamentally different from more problematic kinds of indeterminism in physics (Norton's Dome, quantum indeterminism and diffeomorphism-indeterminism).

Indeed, there is a direct analogy between CTC-indeterminism and the indeterminism arising from lack of knowledge as in statistical physics. In statistical physics data about the macrophysics do not fix the initial conditions of microscopic particles; in the CTC case, initial conditions are hidden from view by Cauchy horizons. Although in statistical physics the microstate is underdetermined, this does not mean that there is no unique future of the system ontologically speaking: according to classical mechanics there is one unique set of initial conditions, one evolution and hence no indeterminism. In the case of CTC-indeterminism, there is also a fact of the matter concerning initial conditions and the precise shapes of the worldlines behind the horizon. Just as we can take away the lack of knowledge and the associated indeterminism in statistical physics by more fully specifying initial conditions, CTC-indeterminism can be taken away by specifying conditions beyond the horizon, namely at the upper side of $L^-$ of Figure \ref{fig:DPST} and at arbitrarily small spheres surrounding the two singular points (points on the lightcones emerging from these points in Figure \ref{fig:DPST}). These conditions added to the initial conditions on $\Sigma$ fully complement the initial-value problem and hence restore global determinism.

Our intuition might suggest that the specification of additional initial conditions on a hypersurface outside $\Sigma$ is unnatural. However, there is nothing in classical mechanics or relativity theory that implies that Cauchy surfaces should have the form of global spatial hypersurfaces. Even in classical theory one can easily define situations in which conditions on a plane cannot fix everything that is going to happen in the future. Think, for example, of a box with a region in it that is shielded from electromagnetic fields. This is a topological structure that in relevant aspects is analogous to that of Figure \ref{fig:DPST}: initial data inside the shielded region will be necessary to achieve a globally deterministic description. Similarly, taking away indeterminism by specifying initial conditions in the region behind the Cauchy horizon is the natural thing to do in time travel situations. 

That there may be a need to explicitly look at the perhaps unusual nature of the global situation in order to determine initial conditions may seem counter-intuitive. However, this feeling is likely due to our impression that we directly experience the world at large, and have access to infinite spacelike hypersurfaces---from which we can make predictions. In reality, however, already special relativity teaches us that we can only have knowledge about limited spacetime regions so that our actual predictions have a local character. In other words, we do not know from experience that there are global Cauchy hypersurfaces and should be open to the possibility that our world is different. From this perspective, CTC-indeterminism is not stranger and more worrisome than the indeterminism arising from lack of knowledge in other areas of physics.

\section{From Indeterminism to Determinism} \label{ItoD}
There is an interesting counterpart to the indeterminism that arises in the presence of CTCs. As we already noted, consistency of histories on CTCs requires that constraints are satisfied that guarantee the uniqueness of events along each CTC. In the case of, e.g., the grandfather paradox these constraints lead to restrictions on what a human agent can do when returning to his own past---this suggests a kind of ``superdeterminism'' that imposes stronger conditions than what we are inclined to expect. The ``lack of freedom'' that results from this is a well-known reason not to take CTCs seriously.

As noted in section \ref{CC}, appeals to human agency and free will can easily obscure the physical points that are at stake. Fot this reason we shall analyze a variation on the grandfather paradox in which a quantum process is considered instead of the actions of a human time traveler. Think of an electron in a quantum state that is a superposition of two different eigenstates of energy, both with equal weights, and with energies $E_1$ and $E_2$, respectively. After some time the electron arrives at a measuring device designed to measure energy and interacts with it. Quantum theory predicts that there is a $50\%$ probability that $E_1$ will be found and recorded, and an equal probability that $E_2$ will be recorded.\footnote{The frequently discussed case of a superposition of \emph{spin} states, e.g.\ the singlet state, is basically identical. It is important to note that the predictions for the outcomes of measurements made by quantum mechanics do not depend on the presence of human observers; quantum predictions are statements about what is recorded by macroscopic measuring devices, regardless of whether a conscious agent becomes aware of the outcomes.}

According to the standard interpretation of quantum mechanics there are no underlying deterministic factors that determine the outcome of this experiment. The quantum state is taken to contain a full specification of everything that is relevant to the prediction, and this state only yields probabilities. In fact, one can derive technical results (Bell's inequalities and their violation) showing that the addition of ``hidden variables'' to the quantum state necessarily leads to theoretical schemes in which the added electron properties behave in unusual and undesirable ways: the hidden properties cannot be purely local (i.e., they must exert an instantaneous influence on each other, even when belonging to physical systems that are far apart). So the indeterminism of quantum mechanics is different from a simple consequence of a lack of knowledge of causal factors. If a proof could be given that quantum indeterminism nevertheless cannot be fundamental, this would be a far-reaching result.

Yet, the consistency conditions that are in force on CTCs seem to imply precisely such a non-fundamentality of quantum mechanical probabilities. Indeed, when we imagine the above-described experiment on a CTC, exactly \emph{one} of the two possible outcomes will be realized---let us assume it is $E_1$. Now, in the future of this outcome we return to the same quantum state of the electron that existed at the beginning of our experiment. Everything will ``repeat itself'', so that the outcome of the experiment will necessarily be $E_1$, the same as before. So there is no question of any probability or uncertainty: the outcome is completely fixed; this is logically forced due to the consistency constraints. Since there is only one unique outcome event on the CTC, with unambiguous properties, we have to conclude that this outcome is determined \emph{tout court}: what is recorded at the end of the experiment is an intrinsically fixed and determined event and there is no place for indeterminism.

If this argument is correct, the feeling that CTCs should be banned from physics by the introduction of a ``chronology protection principle'' may well be justified. Indeterminism as a possible fundamental feature of physical reality would be disproved by a simple CTC thought experiment. This seems to fly in the face of an enormous amount of foundational work in quantum theory. So this version of the grandfather paradox, even though not threatening inconsistency, seems worrisome---at least at first sight.

\section{Determinism versus Determination} \label{detdet}

The argument that CTCs are ``unphysical'' because of CTC-determinism, like the earlier argument from indeterminism, relies on the supposition that we are facing a consequence that only arises in the presence of CTCs. We have argued that this strategy does not work in the case of CTC-\emph{indeterminism}, because this indeterminism also occurs in other situations and is considered harmless there. We will argue now that the case of CTC-\emph{determinism} does not fare better. As a first step we shall show that in many cases CTC-determinism merely exemplifies a kind of logical \emph{determination} that trivially characterizes physical processes in general and does not conflict with \textit{physical} indeterminism.

To see the difference between physical \emph{determinism} and \emph{determination}, think of Minkowski spacetime (Newtonian spacetime will do as well) with a fundamentally indeterministic physical theory defined on it. Physical indeterminism implies that the complete physical state at a certain instant does not completely fix, via the laws of the theory, the physical state at later instants: more than one later states are compatible with the initial state according to the theory in question. Nevertheless, the later states are \emph{determined} in a trivial, logical sense: they are exactly what they are and nothing else. This is just the requirement of unambiguity of events that is needed to make sense of the idea of four-dimensional spacetime at all. If there were no unique physical state of affairs at each spacetime point, we could not have a well-defined history of the universe.

The indispensability of this kind of unambiguous \emph{determination} is especially clear within the conceptual framework of the so-called block universe, associated with the ``B-theory'' of time. According to this B-theory there are no absolute ontological distinctions between Past, Present and Future, and the whole of history can be thought of as laid out in one four-dimensional ``block''. Evidently, all events in this block must be \emph{determined} in the sense just mentioned. It is true that in the literature this has sometimes been taken to entail that the block universe necessarily is subject to determinism. However, this is now generally recognized as a fallacy, at least if \emph{physical determinism} is meant (cf.\ Dieks, 2014, \cite{Dieks2}). Indeed, the block universe exists, according to the B-theory of time, independently of whether the events in it are generated by deterministic laws, of a theory like classical mechanics, or by indeterministic laws like those of quantum mechanics. The crux is that the distinction between physical determinism and physical indeterminism is a distinction between two different kinds of \emph{relations} between physical states: in the deterministic case the laws of the theory make exactly \emph{one} later state compatible with the earlier one, in the indeterministic case there are more later states compatible with the earlier one, according to the theory. By contrast, in the case of logical determination the question of what physical theory applies is not relevant: for an event to be determined in this logical sense it is sufficient that the event is well defined, regardless of its relations to other events. Any \emph{bona fide} unambiguous event is \emph{determined} in this sense.

This argument may seem to rely on the B-theory of time and the associated ontology of the block universe, but on closer inspection it does not. Even if one subscribes to the ``A-theory of time'', and accordingly believes that there are ontological differences between Past, Present and Future, it remains tautological that each future event \emph{will be} exactly and uniquely what it \emph{will be}, and that past events \emph{were} exactly what they \emph{were}.  This tautology merely requires that there is exactly \emph{one} history of the world. So even here, all events are determined in the logical sense, regardless of whether a deterministic or an indeterministic physical theory applies.

Now think back of our thought experiment involving an electron on a CTC. We assumed that indeterministic quantum theory governed what happened in the experiment. This means that the physical state of our electron before its interaction with the measuring device is compatible with different measurement outcomes. This is an assumption about the \emph{relation} between two different states, defined at two different temporal stages of the experiment: the laws of quantum mechanics do not completely fix the measurement outcome when given the  initial electron state as their input.
This relation between the initial and final physical states is objective and unambiguously defined, even on a CTC: when in thought we follow the electron in its history, we return to exactly the same states as before when we arrive at the same points in the electron's existence. So the relation between any two states is uniquely defined. This relation can certainly be the indeterministic one specified by quantum mechanics. Nevertheless, the final outcome of the experiment is unambiguously determined since it corresponds to the single and unique measurement event (that we encounter again and again when we let our mind's eye traverse the CTC various times).

This is fully analogous to what takes place in situations \emph{without} CTCs. It is perhaps easiest to recognize this by again thinking of the block universe: according to quantum theory there are probabilistic relations between the physical properties instantiated at different spacelike hypersurfaces in this universe, but nevertheless these physical properties are unambiguously and uniquely determined in themselves. As we have already mentioned, the same argument can be used even if the notion of a block universe is rejected and some version of the A-theory of time is adopted. Accordingly, if we accepted the principle to brand as \emph{unphysical} processes that are governed by indeterministic physical theories but are nevertheless fully \emph{determined} in the logical sense, we would have to throw out all of physics.

However, this is not the whole story. The plausibility of the argument that CTCs lead to (super)determinism also derives from the fact that \emph{information} about the final outcome may already be present when the experiment starts. For example, we might imagine that the outcome is recorded in a book, which survives along the CTC and can be consulted at the beginning of the experiment. In this case it seems natural to assume that the physical state at the moment that the experiment starts contains this information, so that it becomes possible to derive the experiment's outcome from this initial state. This then appears to lead to physical determinism after all.

In order to discuss this argument and its relevance for the acceptability of CTCs we have to delve a bit deeper into the question of what the physical \emph{laws} on the CTCs look like, and whether these laws enable us to predict the outcome of experiments from their  initial states as just indicated.

\section{Prediction and Retrodiction on CTCs}\label{lawhood}

Suppose that the quantum experiment that we described in the previous section, with possible outcomes $E_1$ and $E_2$, is performed on a CTC and has the actual outcome $E_1$. Suppose further that a record of this is preserved along the CTC, up to the point at which the experiment is about to start. The total physical state at this point on the CTC does not only comprise the state of the electron, but also the record that reveals the outcome of the experiment. Therefore, in spite of the fact that the quantum state of the electron can only provide us with probabilities, the total state makes it possible to predict with certainty what the outcome of the experiment will be.

It should be noted that for this ``deterministic'' prediction the quantum state is not needed at all: the record does not supply a piece of missing information that is needed to supplement what is given by the quantum state, but does all the predictive work on its own. Logically speaking, this is not different from the laboratory situation (without any CTCs) in which a quantum experiment is done, its outcome noted down, after which the record is put in an archive from which we can retrodict the outcome even many years later. This retrodiction does not rely on a quantum mechanical calculation using the later state of the electron (if it still exists), but is based on the classical behavior of records. Books and similar records are designed to be more or less permanent and to follow the deterministic laws of classical physics to a very high degree of approximation. If we only had the later quantum state of the electron to guide us, we would not be able to retrodict unequivocally what the experiment's outcome was; what makes the retrodiction reliable is the approximately classical behavior of the record.
The situation in our thought experiment on the CTC is very similar. It is not the case that quantum theory has ceased to be applicable, but an additional factor has come into play, namely the presence of a classical deterministic record. When the experimenter consults the book, at the beginning of the experiment, and notes the recorded outcome, she is engaged in exactly the same activity as an observer who consults his  notebook in our Minkowski-like world to see what has happened. So just as we are not entitled to conclude that reliable retrodiction falsifies the fundamentally probabilistic nature of quantum mechanics in our world, can we draw that conclusion in the case of the CTC thought experiment.

One might object that the CTC experiment is essentially different from the familiar retrodiction case: what is at stake is not retrodiction at all but rather \emph{prediction} of the outcome of an experiment, on the basis of the initial physical state. But this counterargument does not succeed: the reliability of the statement that the outcome of the experiment will be as the book indicates, derives solely from the book's reliability as a retrodictor, in the same way as in the case without CTCs. The future-directed aspect only comes in because of the uniqueness (and determination) of events along the CTC: as it happens, the retrodicted event is the very same event as the future result of the experiment. So it is the combination of retrodiction and determination that is at work here, and not some new physical principle of prediction.

This becomes more transparent when we ask whether there are reasons to assume the existence of new lawlike physical principles on the CTC. For example, will there be a law telling us about the future? This should not be expected. It is not a general feature of CTCs that there are notebooks or similar records of the eventual outcome available at the beginning of each experiment---these may well have been erased or eroded long ago, if they ever existed at all. There is no reason to assume new regularities on CTCs that ensure that records will be more stable than usual; quite the opposite, if the outcome is to be recorded at the end of an experiment, any already existing record will have to be erased before. It is perfectly consistent to assume the usual laws of physics on CTCs plus the boundary condition of periodicity, i.e. the consistency constaints, without worrying about any new laws.

The consistency conditions that apply in worlds with CTCs thus need not be regarded as symptoms of the existence of novel physical principles, as has sometimes been suggested in the literature.\footnote{For example, Earman (Earman, 1995, \cite{Earman}, p. 194) writes: ``Indeed, the existence of consistency constraints is a strong hint---but nevertheless a hint that most of the literature on time travel has managed to ignore---that it is naive to expect that the laws of a time travel world which is nomologically accessible from our would will be identical with the laws of our world. In some time travel worlds it is plausible that the MRL laws [Mill, Ramsey and Lewis, or the ``best systems account"] include the consistency constraints; in these cases the grandfather paradox has a satisfying resolution. In other cases the status of the consistency constraints remains obscure; in these cases the grandfather paradox leaves a residual itch. Those who wish to scratch the itch further may want to explore other analyses of laws. Indeed, time travel would seem to provide a good testing ground for competing analyses of laws.'' In our view, the consistency constraints merely reflect consistency, which of course has to be satisfied as a trivial logical principle anyway, and do not introduce new laws of physics.} They just reflect uniqueness of events and, indeed, consistency; they have a logical rather than a physical character. There are no fixed patterns of events connected to them. The most one can say is that these consistency conditions enforce a violation of a principle that one is inclined to employ in calculations in which no CTCs are involved, namely that arbitrary initial conditions can be imposed locally---one only needs to think of rolled-up Minkowski spacetime to see that not all such specifications will lead to the periodical solutions that we need to be consistent. 

The idea (which we argue is violated in the CTC-case) that arbitrary initial conditions can be imposed locally has been formulated by Deutsch and Lockwood (Deutsch $\&$ Lockwood, 1994, \cite{Deutsch2}, p.\ 71), which they have called the Autonomy Principle:
\begin{quotation}
\noindent it is possible to create in our immediate environment any configuration of matter that the laws of physics permit locally, without reference to what the rest of the universe may be doing.
\end{quotation}
It is true that \textit{we are not used to constraints on local initial conditions}---but then again, we are not used to taking into account global considerations at all. But, in principle, this is mistaken even in worlds without any CTCs, as we shall note in the Conclusion. At the end of the day, the constraints in question are nothing but the expression of ordinary physical laws in combination with the principles of logic and do not require new principles of physics.

\clearpage
\section{Conclusion}
\textit{Common sense may rule out such excursions---but the laws of physics do not.} 

\hfill ---David Deutsch $\&$ Michael Lockwood (1994, p. 69). \\

The potential existence of closed timelike curves poses a challenge to our intuitions concerning determinism and indeterminism. Physical processes inside a time travel region must satisfy consistency conditions, which constrain the dynamics, and on the other hand the existence of Cauchy horizons implies a lack of predictability. However, when judging the seriousness of this conflict with intuition we should not forget that our common-sense notions about (in)determinism and predictability derive from untutored interpretations of everyday experience. 

One such untutored common-sense idea is that we have immediate access to a global now, a plane, from which the world develops to its future states. Relativity theory has taught us that this notion is mistaken: we do not have epistemic access to a global now at all because of the existence of a finite maximum speed of signal propagation, and ontologically the theory does not single out preferred now-planes. In keeping with this, we should adapt our ideas about causality from  global to local notions. It is a basic message of relativity theory that criteria for determinism should first of all depend on local considerations, and not on considerations about the possible existence of global nows in the universe.

CTC-indeterminism hence turns out to be similar to a familiar and harmless type of indeterminism that occurs in many other places in physics. It is very different from the varieties of philosophically interesting indeterminism that have been focused on in the recent literature, and rather represents another instance of epistemically grounded lack of predictability that is pervasive in physics. From a local point of view, everything is completely deterministic and the lack of global predictability can be rectified by adding extra, not yet considered initial conditions. CTC-indeterminism is therefore not the extraordinary new phenomenon that it sometimes has been suggested to be.

Something similar can be said about the determinism in the sense of ``lack of freedom'' that results from consistency constraints on CTCs. What we are facing here is basically determination of events at the level of logic, instead of physical determinism. Also in this case there is no reason to think that new physical principles, too strange to be true, have to be invoked.

Summing up, both CTC-indeterminism and CTC-determinism do not involve new principles or new laws of physics; although intuitively strange at first sight, analysis shows that they are of the same kind as cases already familiar from well-known and accepted applications of physical theory. As a consequence, there is no justification for the argument that we are here facing phenomena that are so exotic that their potential presence suffices to rule out CTCs. What the \emph{prima facie} implausibility of CTC-determinism and CTC-indeterminism shows is that we have not yet succeeded in adapting our intuitions to the world of general relativity---and that we are unexperienced with traveling through time.

\setstretch{1.0}

\end{document}